\documentclass[a4paper,twoside]{article}
\pdfoutput=1

\usepackage{hyperref}
\usepackage{fancyhdr}
\usepackage{graphicx}
\usepackage{graphicx}
\usepackage{amsmath}
\usepackage{amssymb}

\begin{document}
\author{Oliver D\"urr and Arnd Brandenburg}
\title{Using Community Structure for Complex Network Layout}


\maketitle


\begin{abstract}
We present a new layout algorithm for complex networks that combines a multi-scale approach for community detection with a standard force-directed  design.  Since community detection is computationally cheap, we can exploit the multi-scale approach to  generate network configurations with close-to-minimal energy  very fast.  As a further asset, we can use the knowledge of the community structure to facilitate the interpretation of large networks, for example the network defined by protein-protein interactions.

\end{abstract}



\section{Introduction}

Graph drawing is an important tool in the analysis of networks, which are used in many application areas as diverse as social sciences and bioinformatics.
Simple force-directed layout algorithms can be applied successfully to undirected graphs consisting of a few 100 vertices. However,  for larger graphs,  the standard methods are hampered by very long running times and do often converge to local minima. 

{\it Multilevel} approaches for layouts of large graphs have been successfully used in the past to overcome these limitations. For a recent review, see e.g. \cite{bartel}. A multilevel algorithm usually consists of three steps. In the first step, the {\em coarsening phase}, the graph $G$ is coarsened into a sequence $G_1,G_2,\ldots G_L$ of graphs with decreasing number of vertices. This is done by combining several vertices of the graph $G_i$ into a single vertex of $G_{i+1}$. Then a standard force-directed algorithm is used to layout the smallest graph $G_L$. Finally, in the {\em placement phase}, the positions of the vertices of $G_L$ are used as a starting point for the layout of $G_{L-1}$. In that step, the vertices of $G_{L-1}$ are placed close to their coarsened description in $G_L$. The standard force-directed algorithm is then used again to layout $G_{L-1}$. This procedure is repeated until the original graph $G_1$ is finally layouted.

\par
{\it Communities} in networks are clusters of vertices with a higher than expected fraction of internal edges. Force-directed network layouts that replace edges by pairwise attractive forces between the adjacent vertices can therefore be expected to place vertices of the same community close to each other. Recently Blondel et al.~\cite{blondel} proposed a fast calculation of the community structure of networks which also uses a multilevel approach. 

In this paper we show that the approach of~\cite{blondel} can be used in the coarsening phase of a multilevel layout algorithm.
In Section \ref{sec:method} we give details of our approach, and in Section \ref{sec:results} we compare it with other methods.

\par
In some cases such as the protein-protein interaction (PPI) network, a community structure is expected~\cite{Lewis}, but the layout does not reflect this due to the complexity and size of the network. Our multilevel approach can be generalized naturally by stiffening the strength of the springs inside a community. In this way, the final layout accentuates the decomposition of the network into communities and thus simplifies the interpretation of the result. This will also be shown in Section 3.
\section{Methods}\label{sec:method}

\subsection{Force model and algorithm}

Our layout algorithm uses a standard force-directed method~\cite{Fruchterman91graphdrawing}. It assumes the following three forces: 
a repelling Coulomb-like force ${\bf F}_C$ between all vertices, a spring force ${\bf F}_S$ between the connected vertices, and a drag force ${\bf F}_D$.
The Coulomb-like force between two vertices with charges $Q_i$, $Q_j$  at positions  ${\bf x}_{i}, {\bf x}_{j}$ is given  by

\begin{equation}
 {\bf F}_C = \kappa Q_i Q_j  \frac{ {\bf x}_{i}-{\bf x}_{j}}{|{\bf x}_{i}-{\bf x}_{j}|^3} , 
\end{equation}
where we set $\kappa =1$ and $Q_i=Q_j=3$. The spring forces are given by

\begin{equation}
 {\bf F}_S =  - k S_{ij} \left(|{\bf x}_{i}-{\bf x}_{j}|-r_0\right) \frac{{\bf x}_{i}-{\bf x}_{j}}{|{\bf x}_{i}-{\bf x}_{j}|},  
 \label{eq:spring}
\end{equation} 
where $k=10^{-4}$ is the spring constant, $r_0=50$ is the rest length, and $S_{ij}$ is a factor that can be modified to change the relative strength of the spring force between and within communities. We use $S_{ij}=1$ unless otherwise stated.

To avoid oscillations, a drag force acts on a vertex moving with velocity $\dot {\bf x}$  according to
\begin{equation}
{\bf  F}_D =   - c \dot{{\bf x}},
\end{equation} 
with $c=0.01$.
The many-particle force ${\bf F}_C$ is efficiently handled by a Barnes-Hut approximation~\cite{BH}, and the equations of motion are solved via Runge-Kutta integration.

\subsection{Multilevel Approach}
The detection of the community structure can be achieved by maximizing the {\it modularity} $Q$, which is defined as
\begin{equation}
Q=\frac{1}{2m}\sum_{i,j}\left(A_{ij}-\gamma\frac{k_ik_j}{2m}\right)\delta(C_i,C_j),
\end{equation}
where the sum runs over all pairs of vertices. The adjacency matrix element $A_{ij}$ represents the weight of the edge between the vertices $i$ and $j$,  $k_i=\sum_j A_{ij}$ is the sum of the weights of the edges attached to vertex $i$, $m=1/2\sum_{ij}A_{ij}$,  and  $C_i$ denotes the community of vertex $i$. The so-called  resolution parameter $\gamma$ was introduced in~\cite{reichardt}, where it was shown that maximizing the modularity is equivalent to finding the ground state of a spin glass model.
It has been shown that maximizing the modularity is NP-complete~\cite{brandes2008}. A heuristic which is able to efficiently discover the community structure in graphs consisting of several millions of vertices has been proposed recently~\cite{blondel}. This algorithm generates successive coarsenings of the original graph until it reaches a final level where the modularity cannot be optimized any further. 

The algorithm starts with the original graph, where each vertex forms a community of its own and all edges have weights equal to one. Then neighbors are placed in the same community if that increases the modularity. This is repeated until no gain in the modularity can be observed. Now all vertices which belong to the same community are joined into a {\it meta-vertex}. While the vertices of the original graph have weight zero, nonzero weights are assigned to  the meta-vertices in the subsequent levels.
The weight of these meta-vertices is given by the sum of the weights of the previous graph plus twice the sum of the edge weights inside the community. The edge weights between two (connected) meta-vertices are simply the sum of the edge weights connecting them. The same approach as for the original vertices is now applied to the meta-vertices. This procedure is repeated iteratively until no gain in the modularity can be observed anymore. The final meta-vertices are the communities. In the following we number the levels of aggregation with an index $i$, where $i=1$ denotes the first level, corresponding to the original graph, and $i=L$ denotes the final level.

We exploit the multilevel structure of the community detection for the graph layout by starting at the coarsed level $i=L$, in which the graph has the same number of vertices as there are communities.
The spring constants at arbitrary levels are given by the weights of the meta-edge times the original spring constant, and the charge of a meta-vertex is its weight times the standard charge. To layout the graph $G_L$, we start with a random configuration 
and perform $n_L$ Runge-Kutta steps. 

There are several possibilities to determine the starting positions of the vertices in the $(i-1)$th level, given the final configuration in the $i$th level,  see~\cite{bartel} for some examples of this placement step. 
Here we determine the initial configuration of the $(i-1)$th level by randomly placing all vertices inside a circle around the corresponding vertex from level $i$. The radius of this circle is taken to be half of the minimum distance to all other vertices.
The above steps are repeated until the final level $i=1$ is reached. 

\begin{figure}[h]
\begin{center}$
\begin{array}{cccc}
	\includegraphics[width=.22\textwidth]{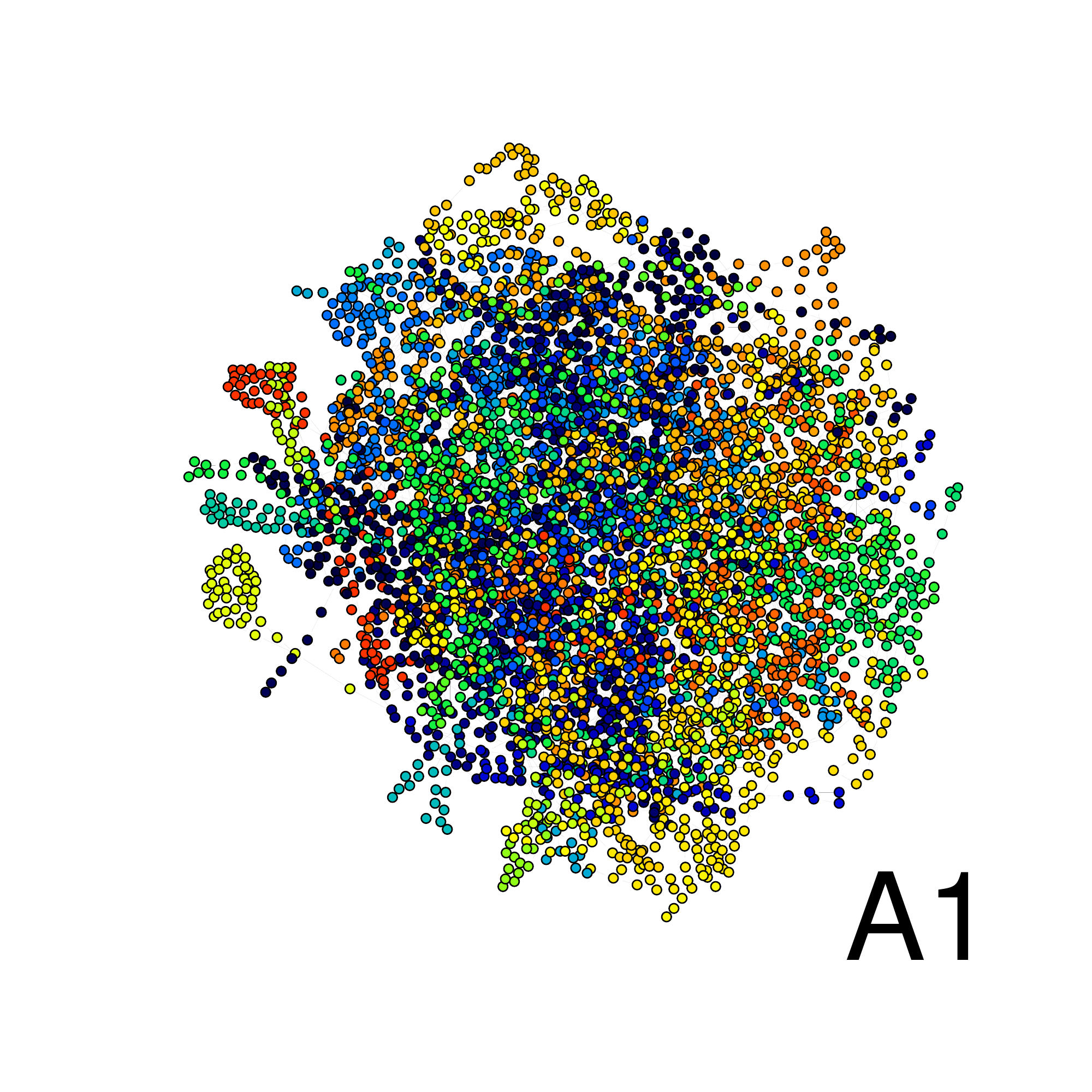} &
	\includegraphics[width=.22\textwidth]{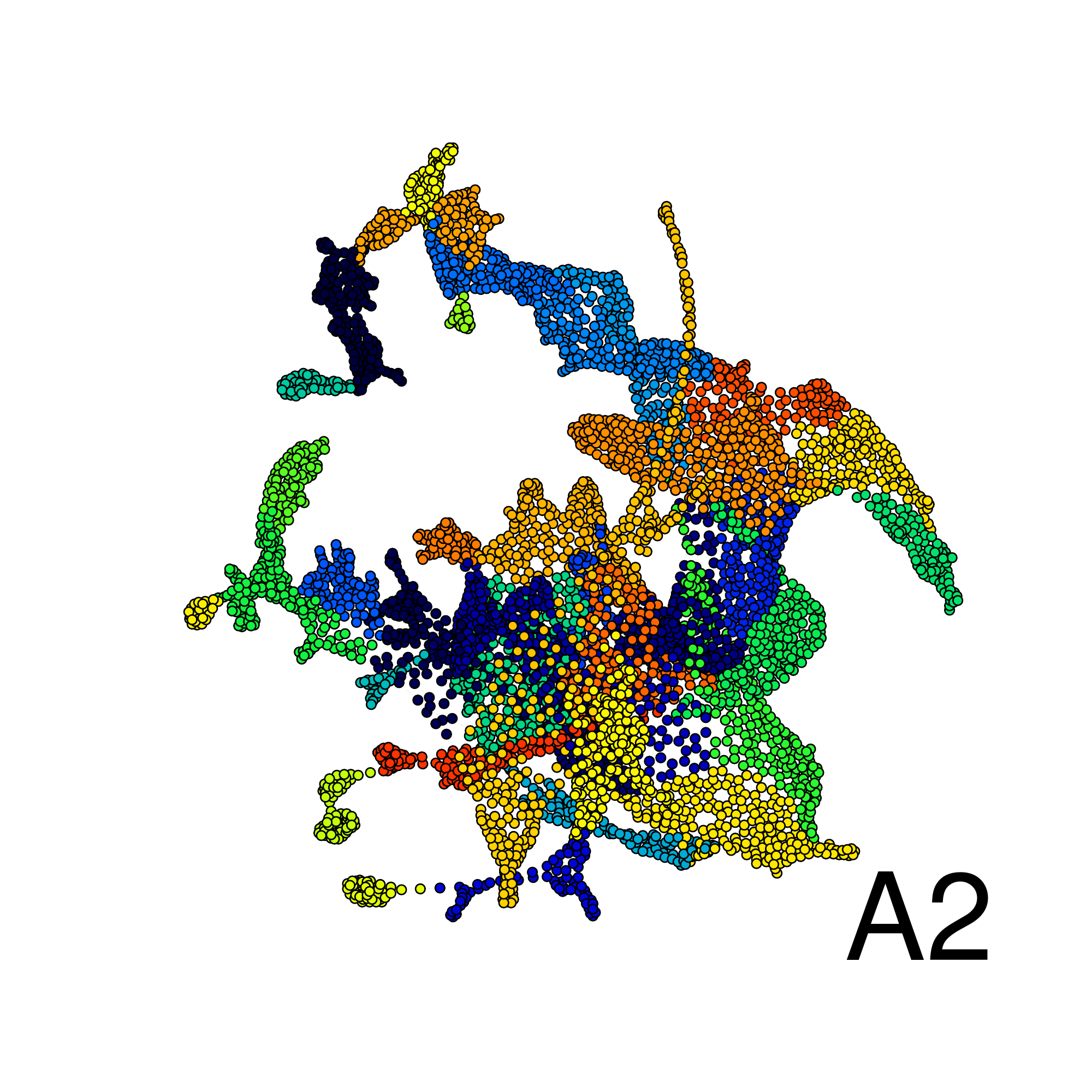} &
	\includegraphics[width=.22\textwidth]{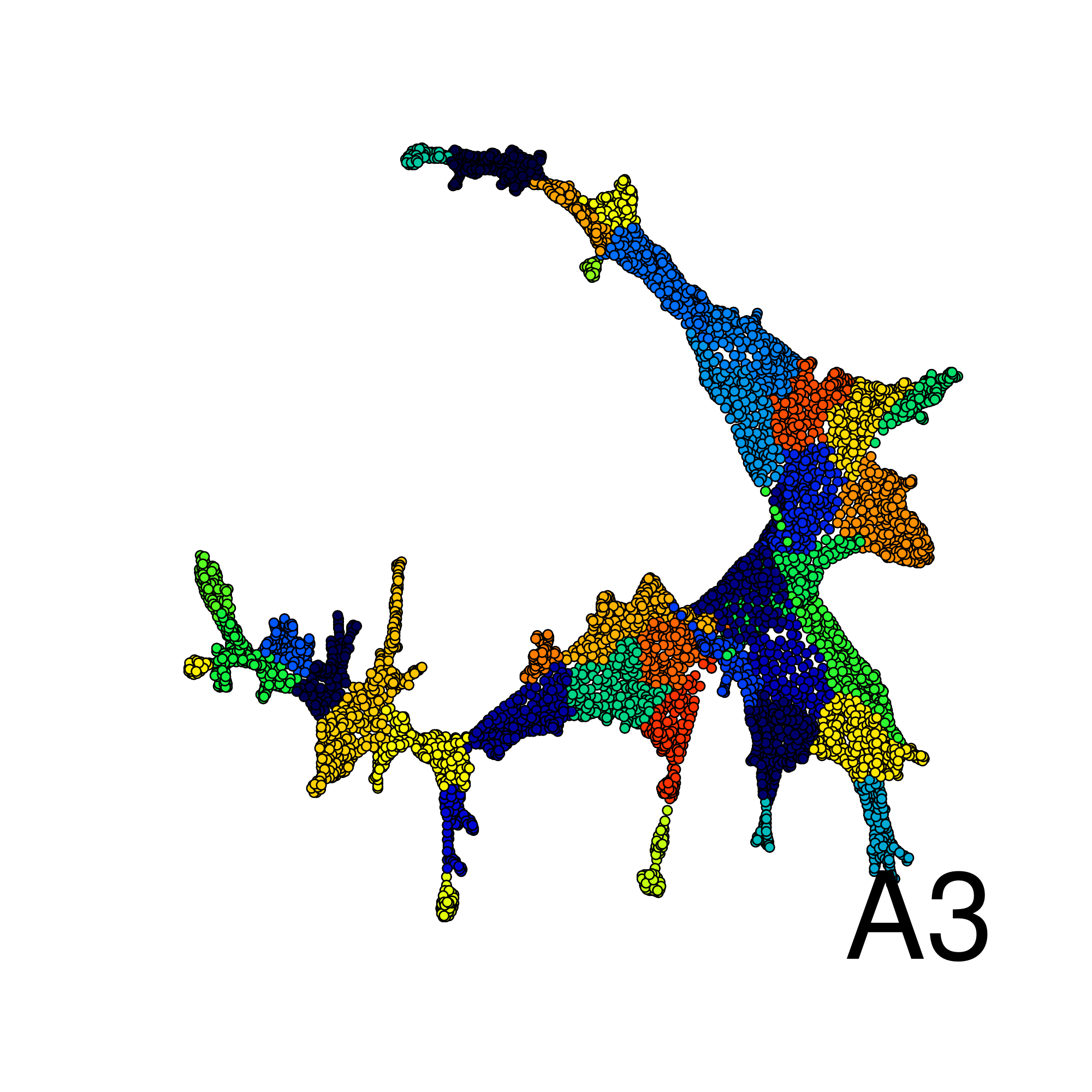} &
	\includegraphics[width=.22\textwidth]{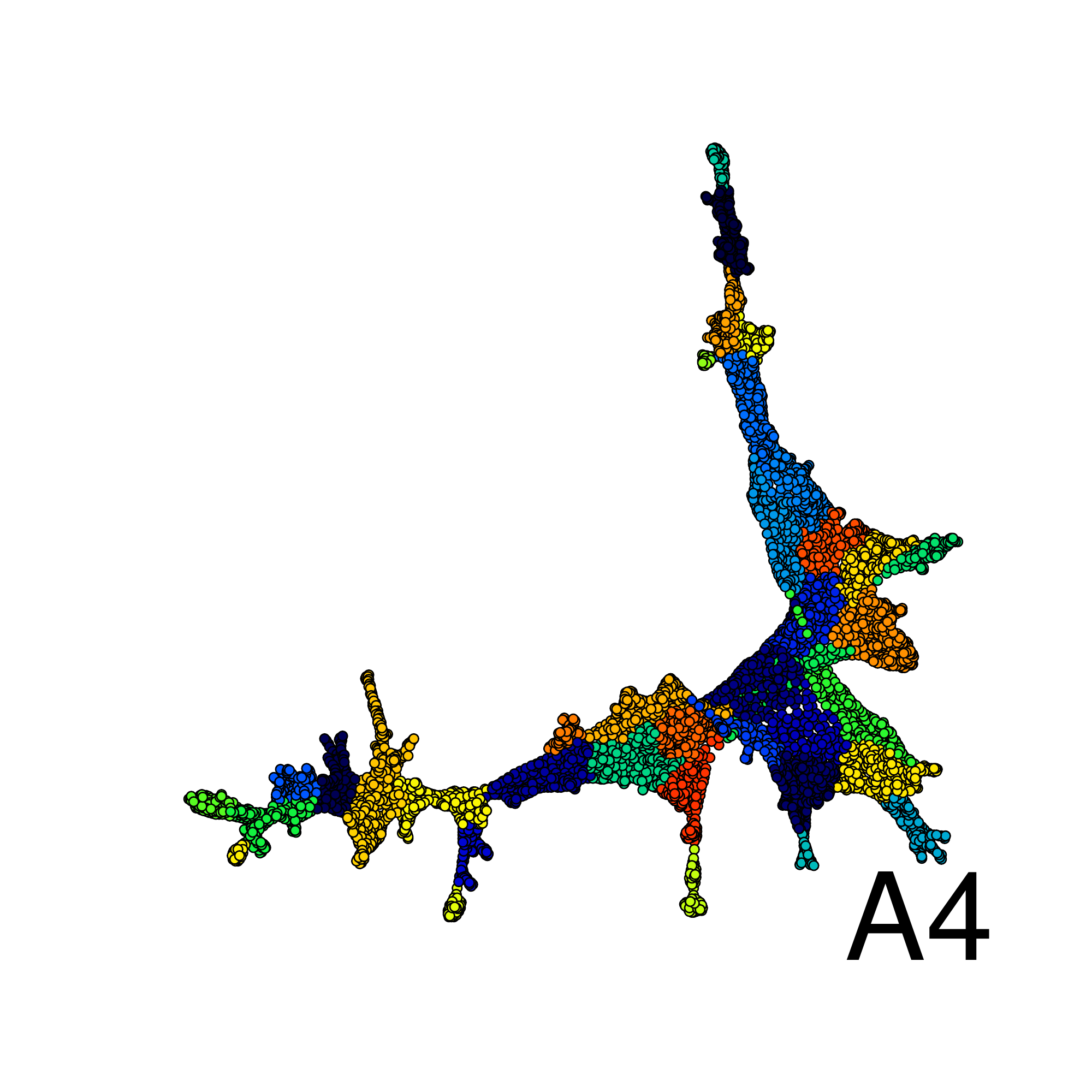} \\
	\includegraphics[width=.22\textwidth]{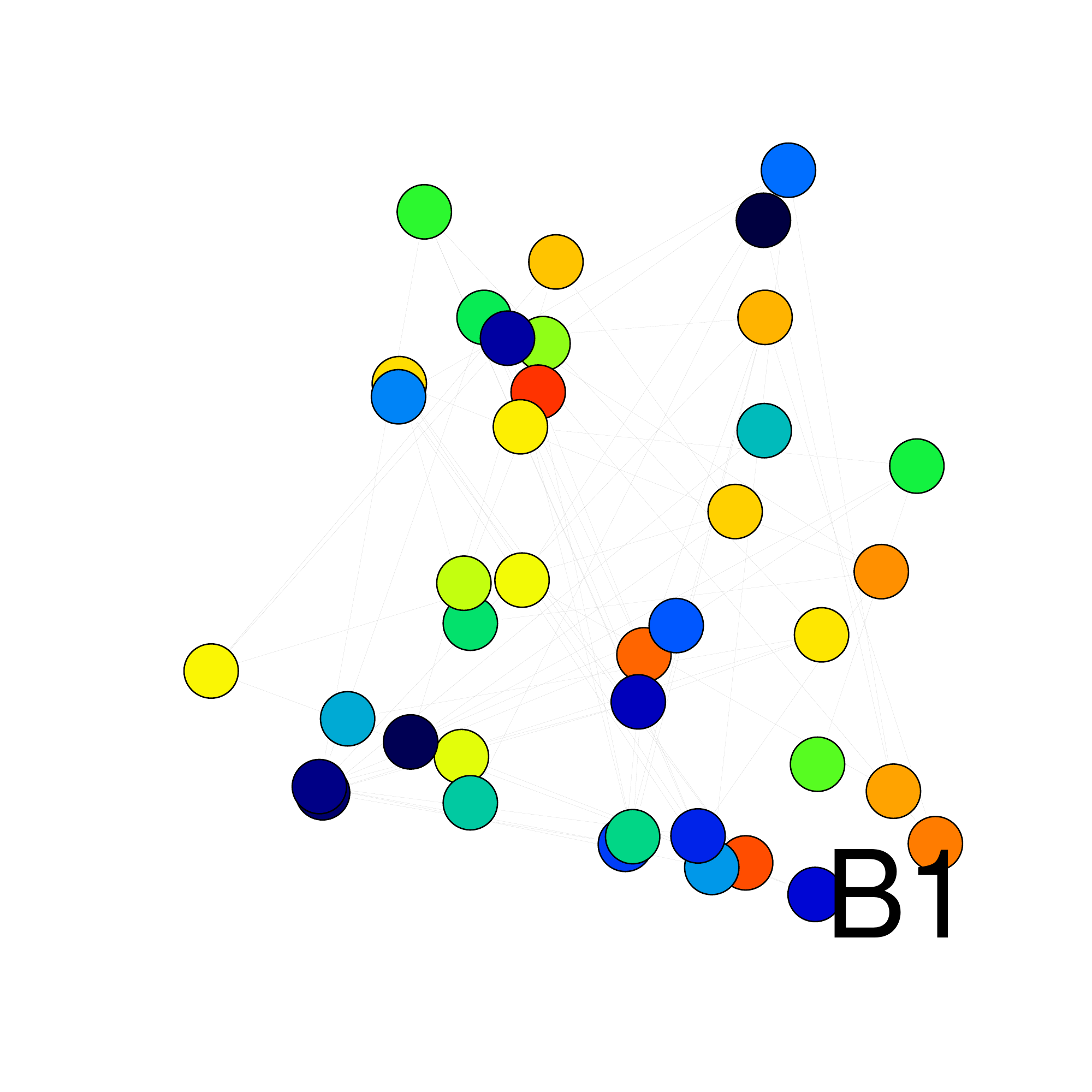} &
	\includegraphics[width=.22\textwidth]{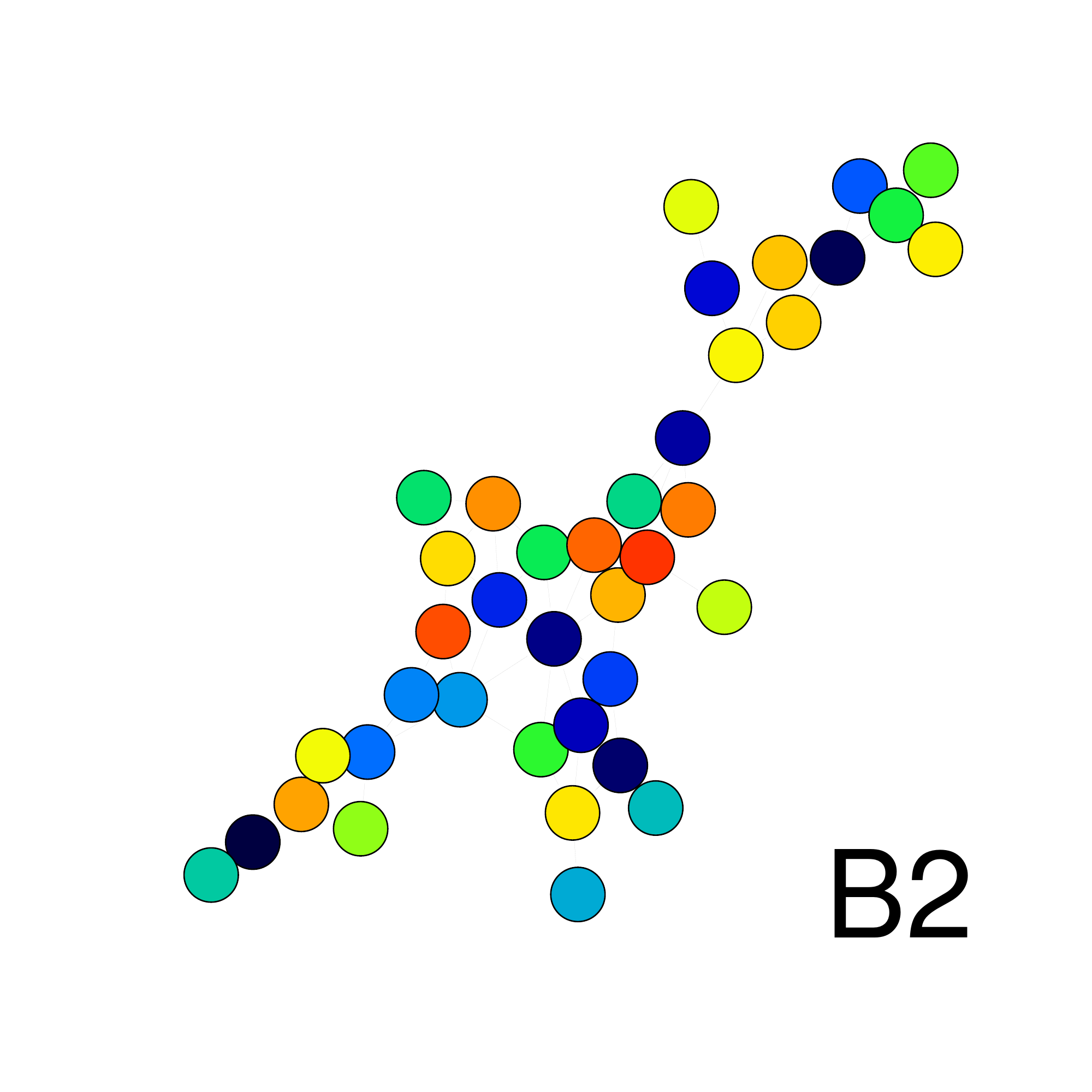} &
	\includegraphics[width=.22\textwidth]{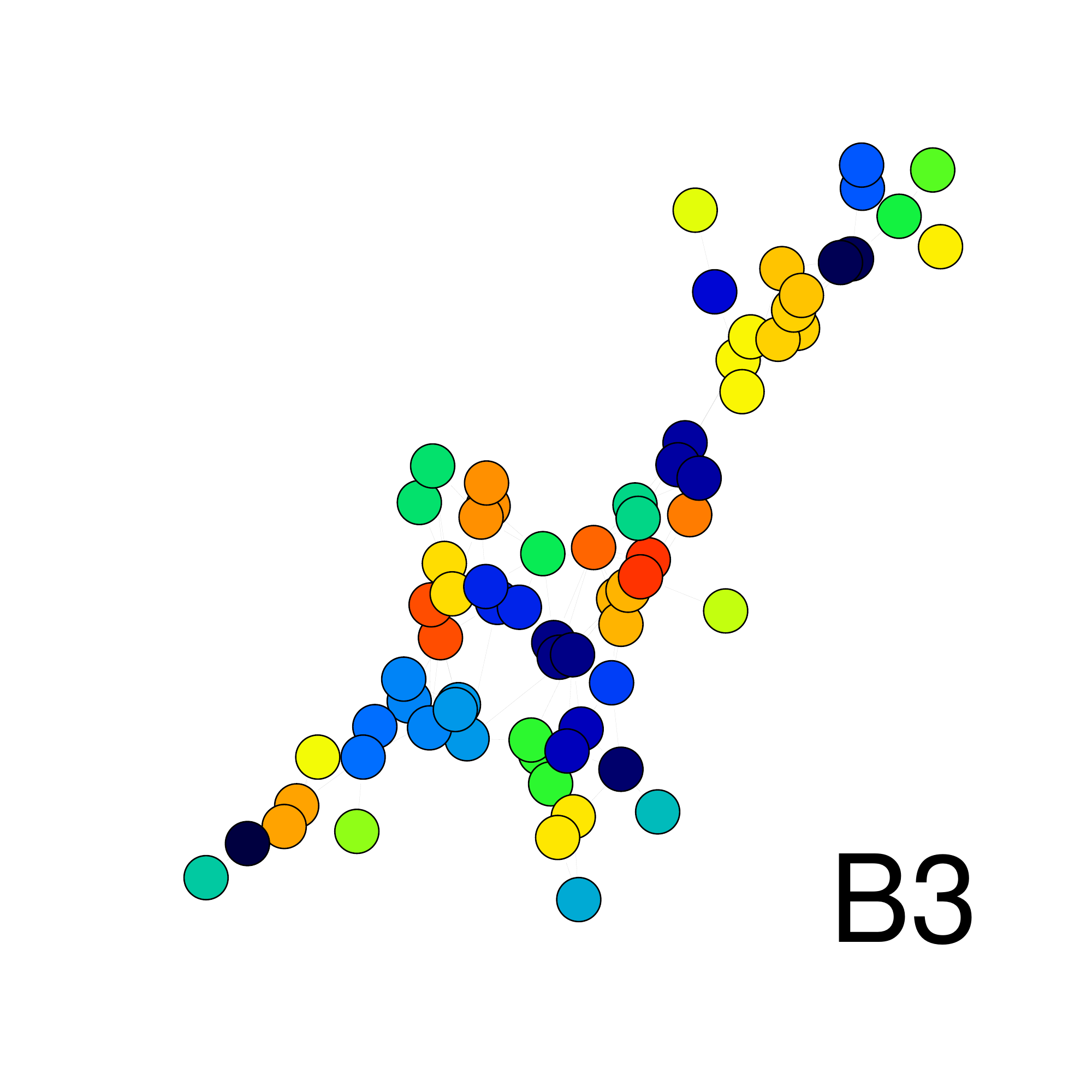} &
	\includegraphics[width=.22\textwidth]{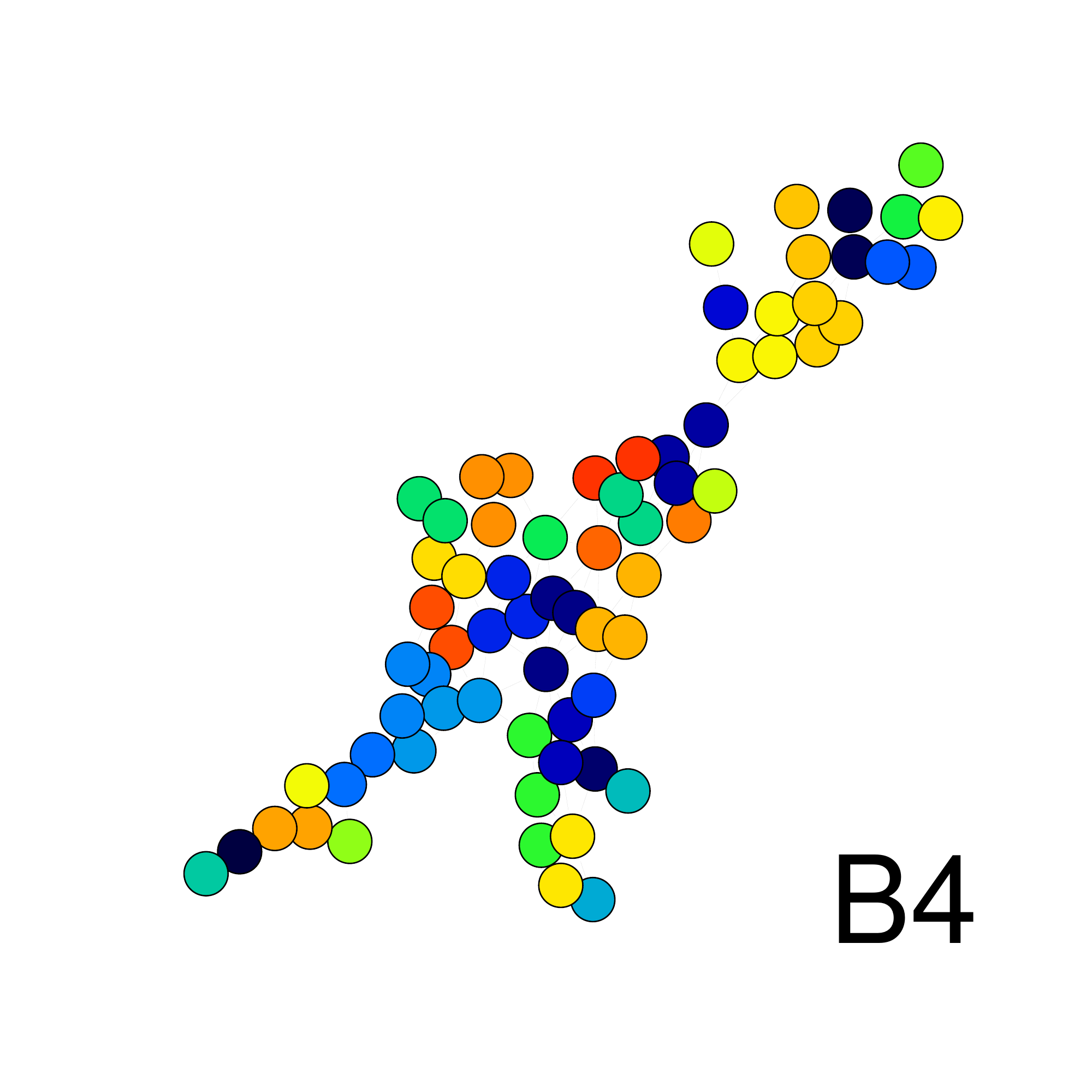} \\
	\includegraphics[width=.22\textwidth]{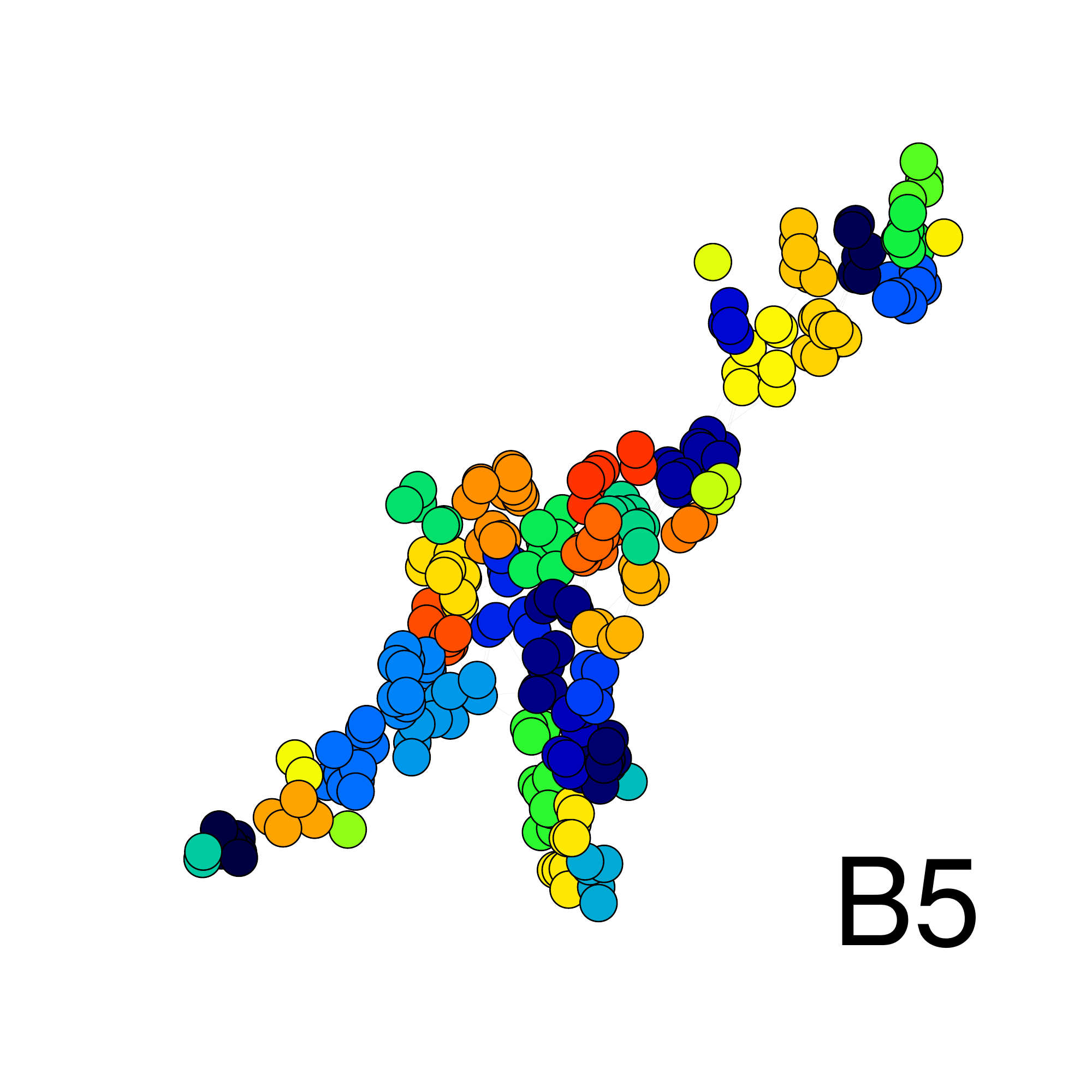} &
	\includegraphics[width=.22\textwidth]{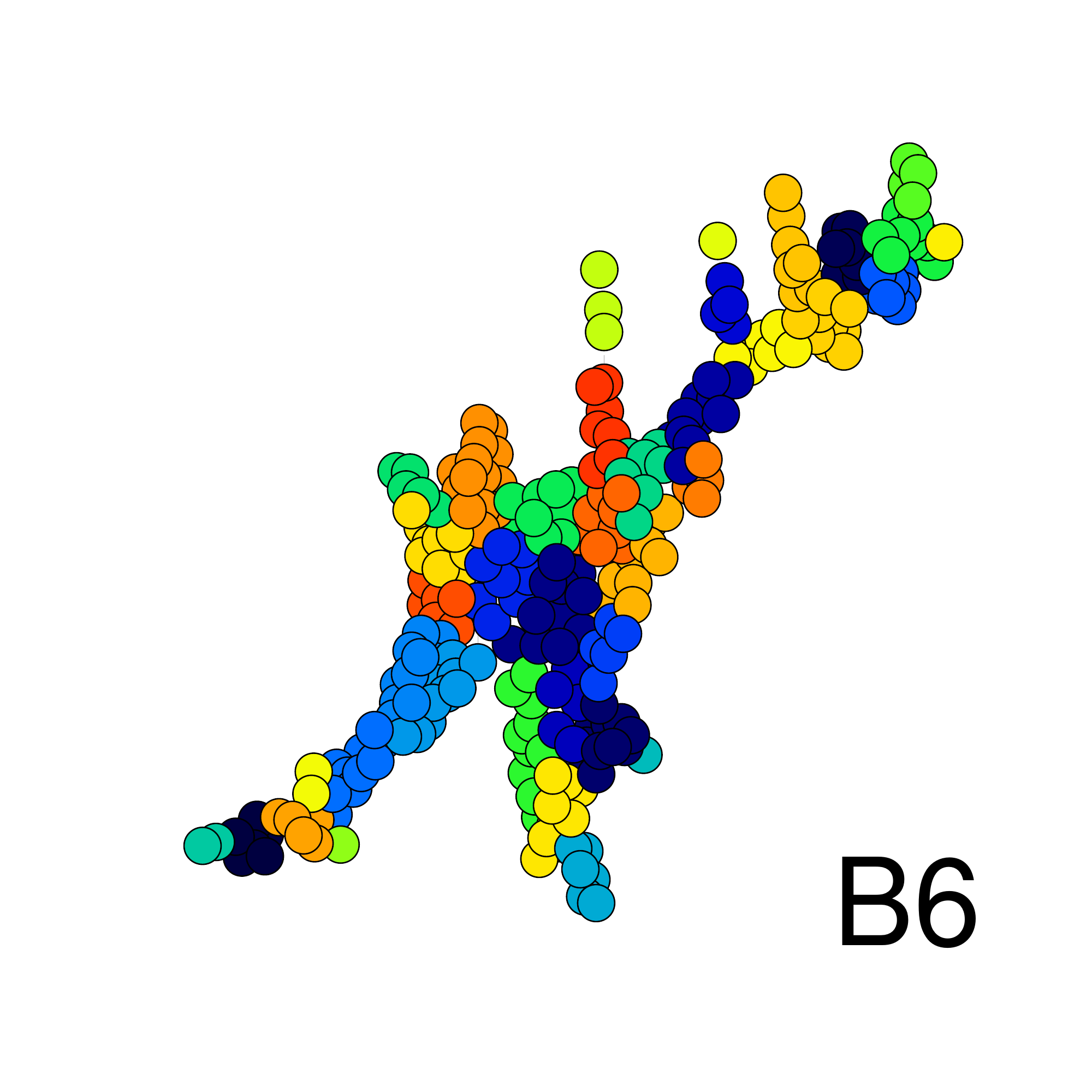} &
	\includegraphics[width=.22\textwidth]{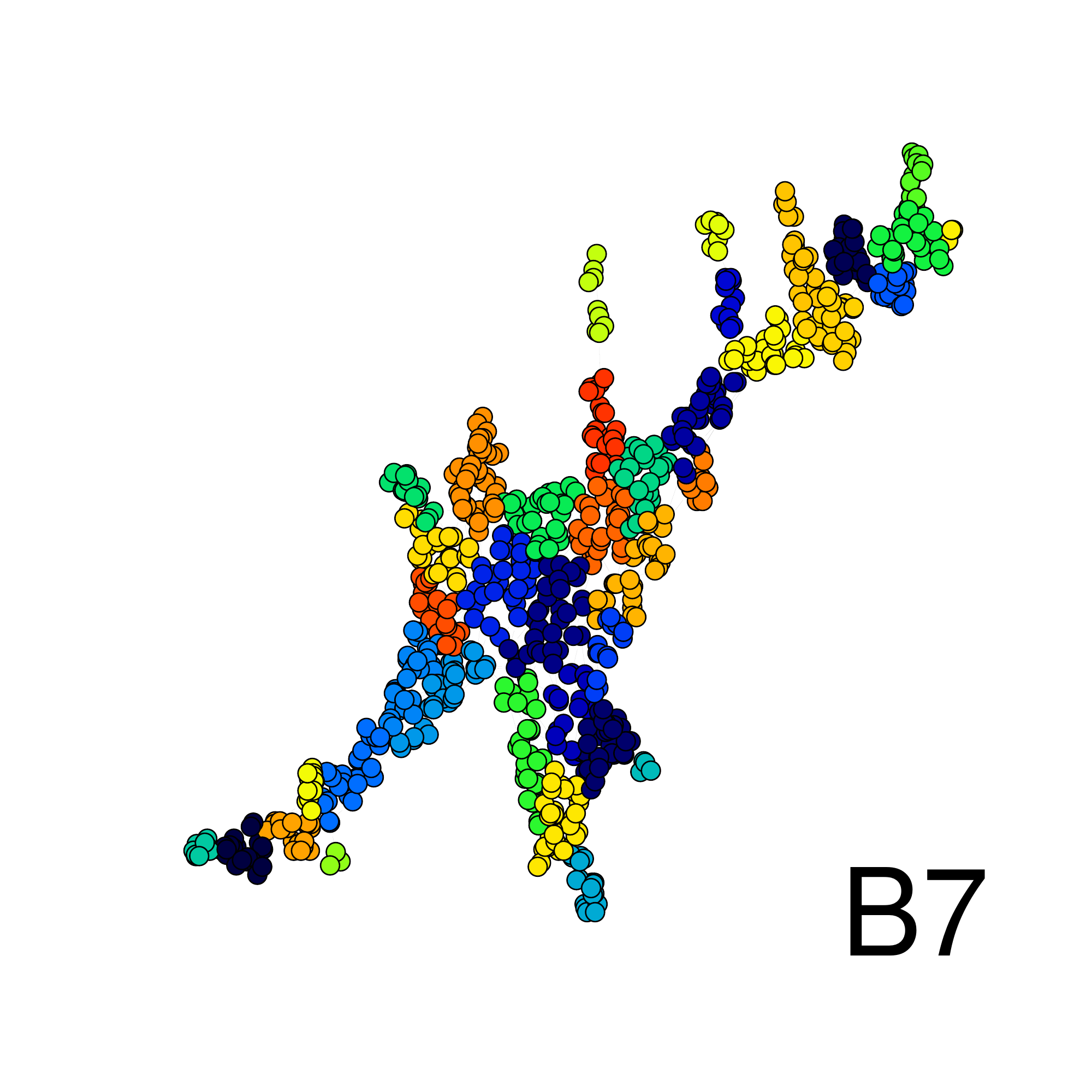} &
	\includegraphics[width=.22\textwidth]{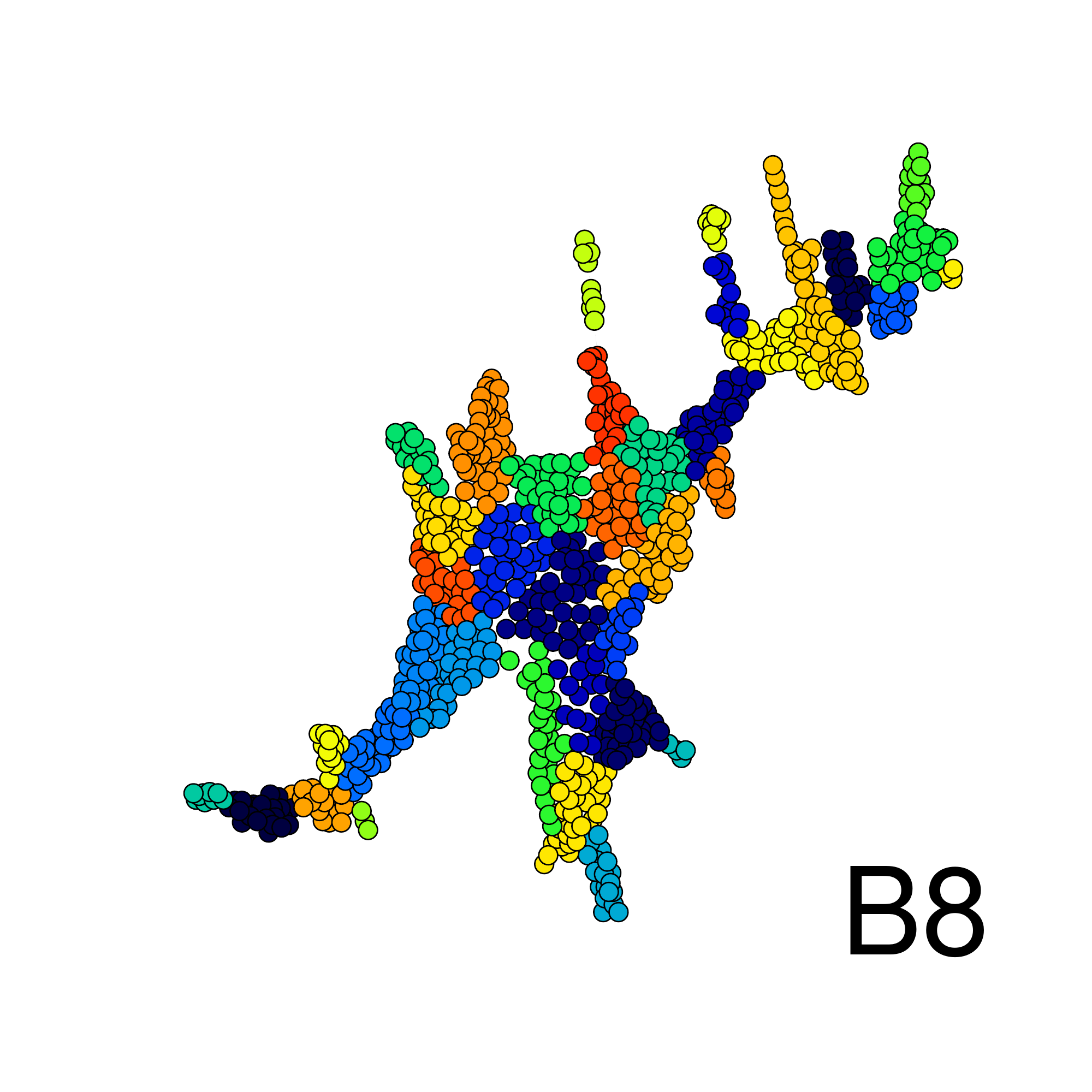} \\
	\includegraphics[width=.22\textwidth]{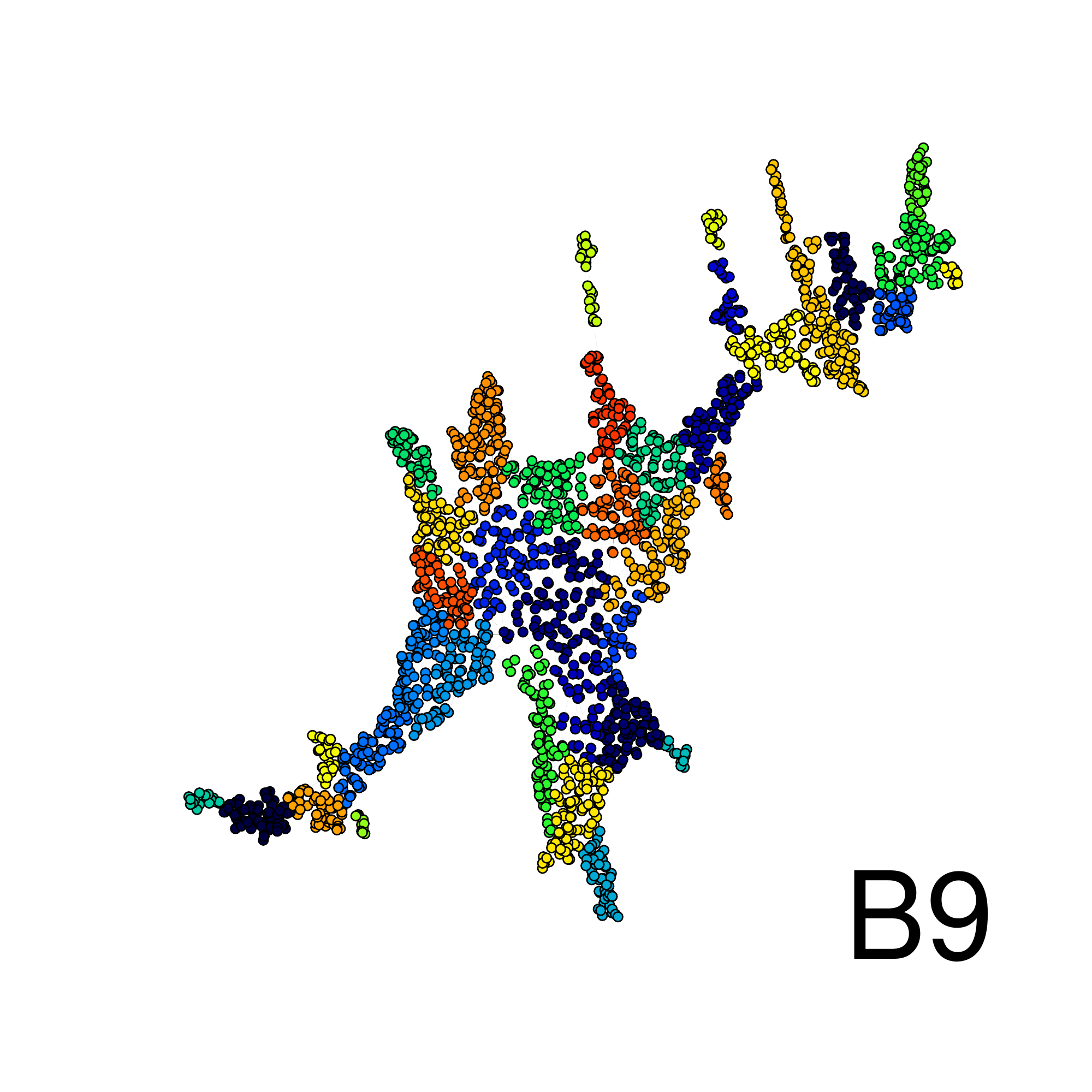} &
	\includegraphics[width=.22\textwidth]{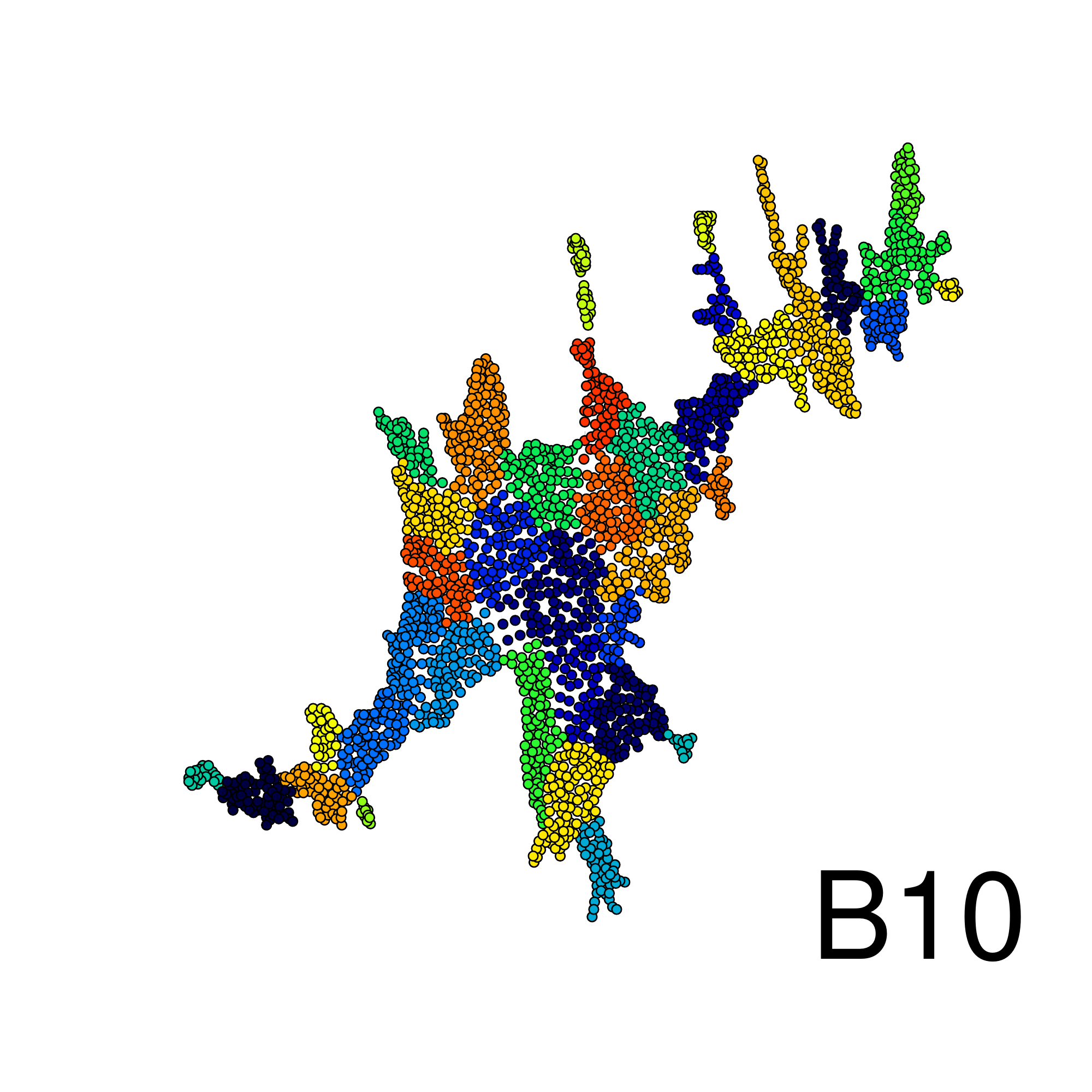} &
	\includegraphics[width=.22\textwidth]{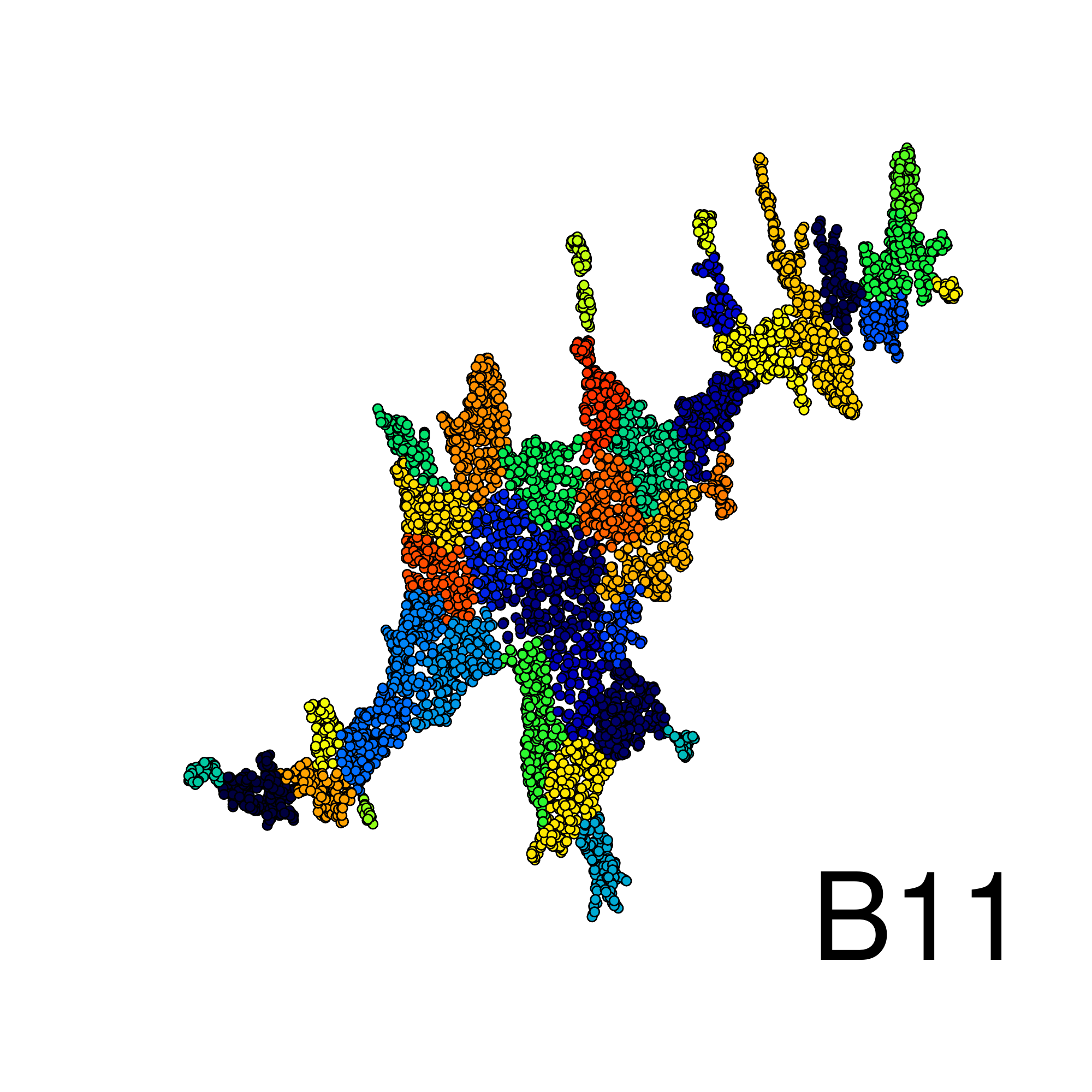} &
	\includegraphics[width=.22\textwidth]{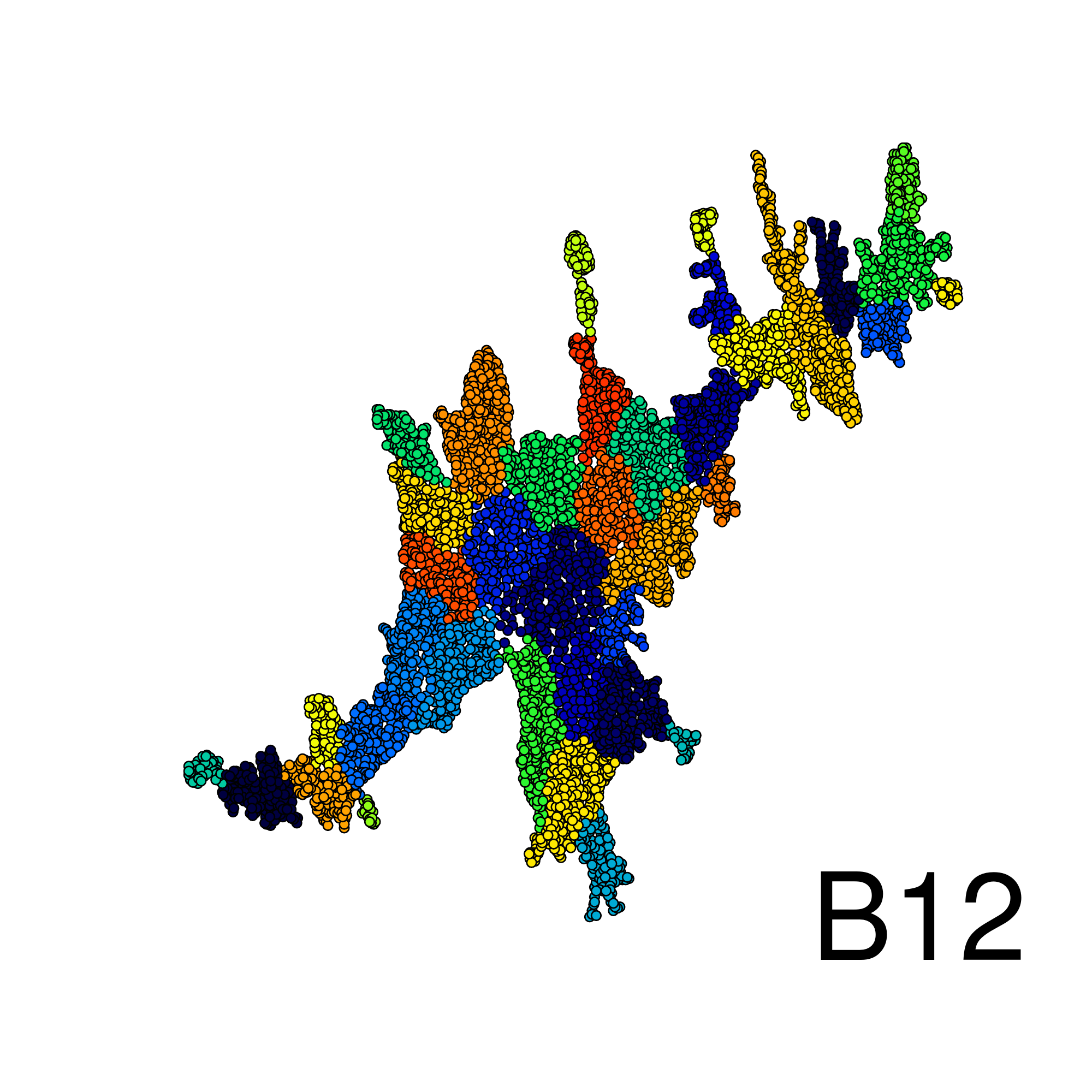} \\
\end{array}$
\end{center}
\caption{The layout for the UK graph using a standard force-directed method (upper row, tiles A1 to A4) and our new multi-scale algorithm (lower rows, tiles B1 to B12). The graph is colored according to the 39 communities detected. In the graphs generated by the standard method, the number of Runge-Kutta steps are $n=100$, $n=1006$, $n=9970$ and $n=393677$ for A1, A2, A3 and A4 respectively, corresponding to CPU times of 2.9, 31, 320, and 12203 seconds. In the multi-level approach,  tiles B1$-$B12, the number of steps for each level is determined via Equation (\ref{eq:ml}) with $n=100$, and the total CPU time was 3.1 seconds. Videos showing the progress of the layout procedure are available at~\cite{DuerrForce} and~\cite{DuerrML}.}
\label{fig:uk}
\end{figure}

To determine the number of Runge-Kutta integration steps $n_i$ in the $i$th level we require that the same amount of computation time is spent on each level. The performance of the algorithm is limited by the calculation of the many-particle forces. The Barnes-Hut treatment of the vertices in the $i$th-level scales like $|V_i| \ln (|V_i|)$~\cite{BH}, where $|V_i|$ denotes the number of (meta-)vertices in level $i$. Hence, we set 
\begin{equation}
  n_i = \frac{n}{L} \; \frac{|V_1| \ln(|V_1|)}{|V_i| \ln(|V_i|)},
  \label{eq:ml}
\end{equation} 
where $n=\sum_i n_i$ is the total number of Runge-Kutta steps.
This also enables an easy comparison of the multiscale approach with non-multiscale ($L=1$) algorithms using $n$ integration steps. 

\section{Results\label{sec:results}}

In Figure \ref{fig:uk} we show results of applying our approach to the so-called UK dataset with $|V|=4824$ vertices and $|E|=6827$ edges taken from~\cite{walshaw}. Videos of the unfolding of the graph with increasing number of integration steps are also available at~\cite{DuerrForce} for the standard force-directed layout and at~\cite{DuerrML} for the multi-level approach. 

\subsection{Comparison with a standard force-directed approach}
In the upper row of Figure \ref{fig:uk} (labeled A1 to A4), the graph layout obtained with the standard force-directed method is displayed after $n=100$, $n=1006$, $n=9970$ and $n=393677$ integration steps.
In the lower rows (labeled B1 to B12) the various steps of our multiscale approach are illustrated. The community detection algorithm determines $L=6$ levels and finds 39 communities in the data set. The starting point of the multiscale algorithm for the graph layout is therefore a random configuration of 39 meta-vertices, each one corresponding to a community. This is shown in tile B1. After 4772 integration steps, the configuration shown in tile B2 is reached. Note that the global structure is already visible in that early phase of the algorithm. In tile B3, some of meta-vertices are replaced by the correponding meta-vertices at level $L=5$. This configuration ($|V_5|=71$, $|E_5=141|$) is further evolved with $n_5=2253$ steps (see tile B4). The remaining tiles B5-B10 correspond to the levels $i=4,3,2$ with number of integration steps $n_i = 461, 121, 38$ and corresponding number of vertices $|V_i| =  265, 836, 2281$. Tile B11 is the starting configuration for the final $n_1=16$ integration steps. Note that the total computational time needed for the final layout of the multiscale approach corresponds to the time needed to get the layout displayed in tile A1 with the standard force-directed method, where no global structure is visible. The total computation time was 3.1 seconds, while the time for detecting the communities was 72 milliseconds and can thus be neglected. 

\par
\begin{figure}[h]
\centering
\includegraphics[width=.95\textwidth]{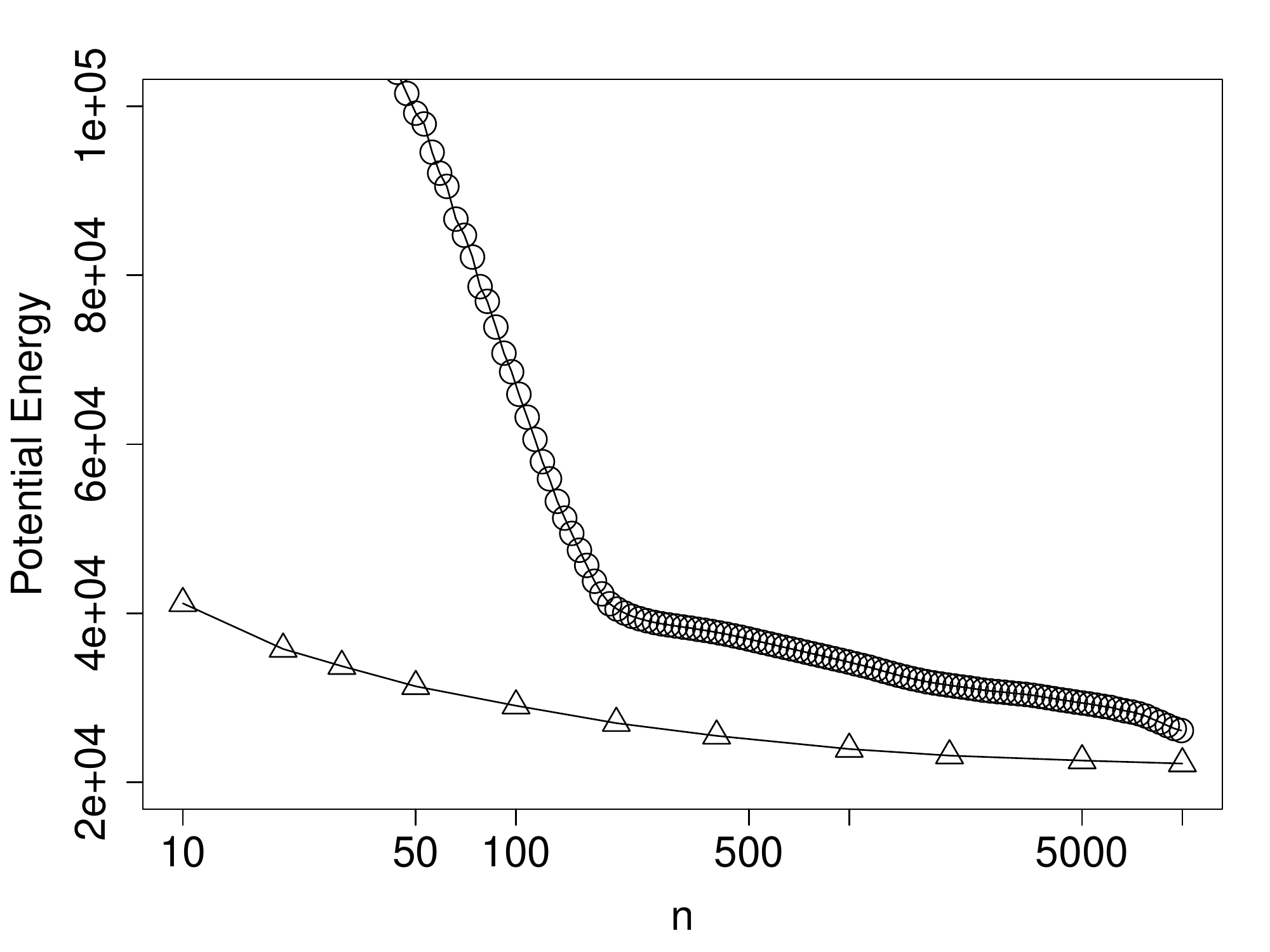}
\caption{Comparison of the energies (in arbitrary units) of the UK graph layout between a standard force-directed layout (circles) and the multiscale approach (triangles) as a function of the total number of integration steps $n$.}
\label{fig:ukenergy}%
\end{figure}

 To compare the algorithms quantitatively, we computed the total energy of the configuration of the UK graph for various total number of integration steps $n$ (see Figure \ref{fig:ukenergy}) for the standard force-directed layout and our multiscale approach, where $n$ is the same for both methods and is distributed between the different levels of the multiscale approach according to Equation (\ref{eq:ml}). For $n=10$, the multiscale approach generates a configuration with an energy that is reached with  the standard method only after about 200 steps. The reason is that the communities and corresponding meta-vertices are determining a close-to minimal energy configuration of the UK graph already at the coarsest level, so that no major, long-range layout changes are needed. This in turn is due to the fact that communities are clusters of vertices with many edges (springs), and therefore keeping vertices of the same community close together is favorable to the energy minimization.

\subsection{Comparison with other multi-level approaches}
\begin{figure}[h]
\centering
\includegraphics[width=.95\textwidth]{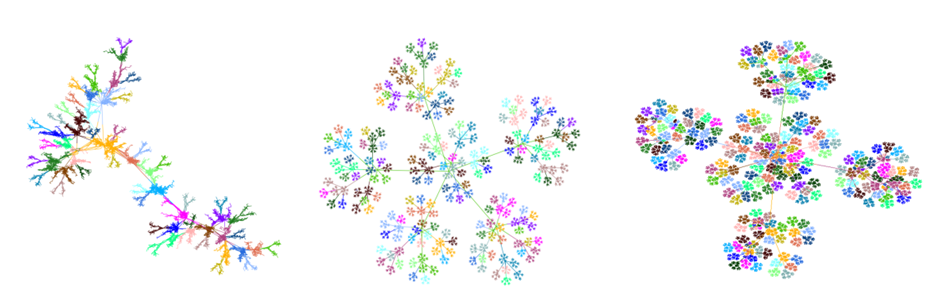}
\caption{Final layouts of data sets used for the comparison in Table \ref{tab:comp} and generated with our method. From left to right: add32, tree\_06\_05, and tree\_06\_06.}
\label{fig:termalization}%
\end{figure}

There are several layout algorithms using a multiscale ansatz~\cite{bartel}. While a thorough comparison is beyond the scope of this work, we compare the running time of our method with two recent implementations of the multiscale FM3 algorithm. In Table \ref{tab:comp}, the running times of our Java implementation on a standard quad core desktop PC are compared (where availible) for some data sets against a standard CPU-based implementation~\cite{Hochul} and a highly sophisticated implementation using  graphic cards processors (GPU)~\cite{Godiyal}. The authors of the GPU-based approach claim that their approach is about $20-60$ times faster than existing CPU implementations~\cite{Godiyal}. 
\begin{table}\label{tab:comp}
\begin{tabular}{| l r|| r | r |r|}
  \hline   
  Data Set &($|V|$; $|E|$)		& FM3 \cite{Hochul} & FM3 (GPU)  \cite{Godiyal} & Our method \\ \hline                    
  add32	&(4960; 9462)& 12.1 & 1.4 & 0.7 \\ \hline
  uk	&(4824; 6837) & - & - & 3.1 \\ \hline
  tree\_06\_05 	&(9331; 9330) & 17.7 & - & 14.1  \\ \hline
  tree\_06\_06 	&(55987; 55986) & 121.3 & 24.6 & 93 \\ \hline
\end{tabular}
\caption{Comparison of running times in seconds for several data sets.}
\end{table}
For the add32 dataset we choose $n=20$, while for all other datasets we set $n=100$. The final configuration for the UK data set is shown in tile B12 of Figure \ref{fig:uk}, the layouts for the other data sets are displayed in Figure \ref{fig:termalization}.

\subsection{Community-aware layout}
\begin{figure}[h]
\centering
\includegraphics[scale=0.55]{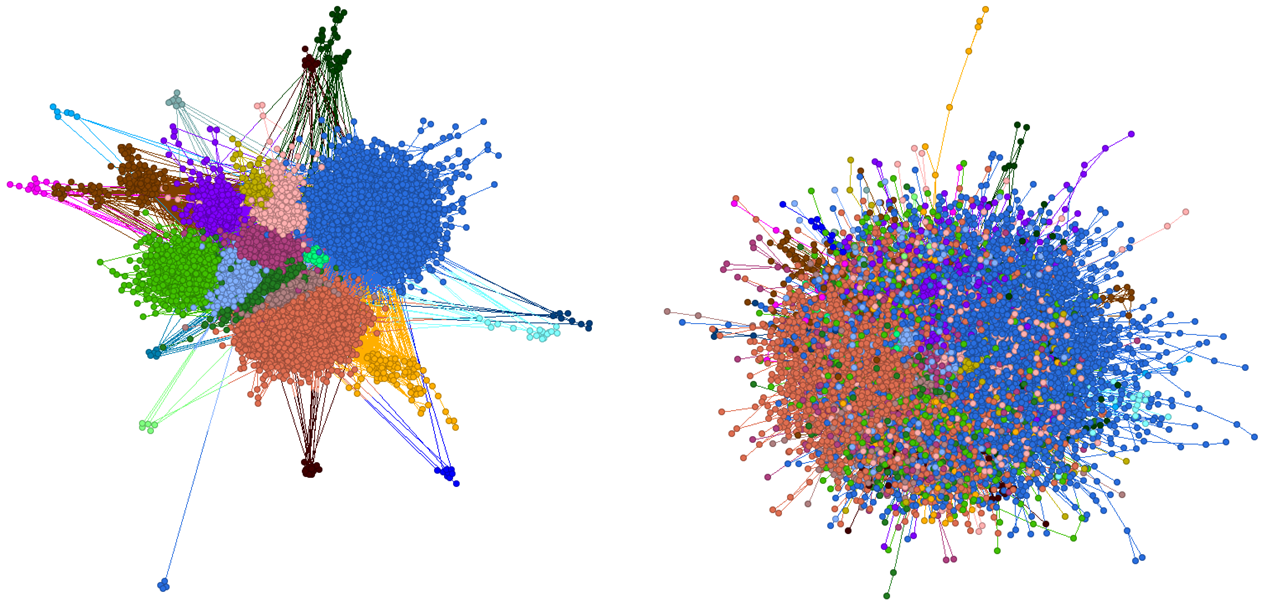}
\caption{Left: Protein interaction network with layout computed using stiffened springs within communities, $S_{ij}=100$. Right: Same network, but using the standard spring strength for all edges, $S_{ij}=1$.}
\label{fig:ppi}%
\end{figure}

Sometimes networks are so entangled that the community structure is not evident in the layout anymore. As an example we consider the human protein interaction network downloaded in June 2011 from the ConsensusPathDB~\cite{cpdb}. The network has $|V| \approx 13000$ vertices and $|E| \approx 240000$ edges.
This network is shown  on the right hand side of Figure \ref{fig:ppi}. One would expect to see a community structure in the network, but the layout does not even partially reflect the community structure due to its complexity and size. Further, it is known that functionally related proteins preferably cluster in the same community~\cite{Lewis}. Hence, also for the visual inspection of the network it is beneficial if communities are layouted in distinct regions of the graph. To achieve this, we set the constant $S_{ij}$ in the spring force Equation (\ref{eq:spring}) to $S_{ij} = 100$ for edges belonging to the same community, and to $S_{ij}=1$ for edges connecting different communities. The resulting network is shown on the left hand side of Figure \ref{fig:ppi}. The community structure is clearly exhibited.
\section{Conclusions}
Generating adequate and useful layouts for complex networks is a challenging task. Utilizing recent progress in the detection of communities in networks, we constructed a new, multi-level layout algorithm that generates configurations of close-to-minimal energy very fast. For the examples studied in this paper, our method outperforms the (standard implementation of the) FM3 algorithm. We expect that our method works particularly  well whenever a clear community structure is present in a network. Knowledge of the communities can also be used to create layouts which accentuate these structures, thus making it easier to understand complex relationships within large networks.

\section*{Acknowledgements}
We would  like to thank J. Hoefkens for carefully reading the manuscript, C. Walshaw for information about the UK graph and U. Brandes for pointing out relevant literature.

\clearpage

\bibliography{Duerr_Brandenburg_7_11_2012}
\bibliographystyle{abbrv}

\end{document}